\documentclass[1p]{elsarticle}
\usepackage{amsmath}
\usepackage{amssymb}
\usepackage[colorlinks=true,urlcolor=blue,citecolor=blue,linkcolor=blue]{hyperref}
\usepackage{mathfunc}

\def\arxivid{1209.3971}
\def\eprint{arXiv:\href{http://arxiv.org/abs/\arxivid}{\arxivid}}

\journal{Computer Physics Communications}
\biboptions{numbers,compress}

\begin{document}
\begin{frontmatter}
\title{NumExp: Numerical epsilon expansion of hypergeometric functions}
\author{Zhi-Wei Huang}
\ead{zwhuang@whu.edu.cn}
\author{Jueping Liu\corref{corauthor}}
\ead{jpliu@whu.edu.cn}
\cortext[corauthor]{Corresponding author.}
\address{College of Physics and Technology, Wuhan University, Wuhan 430072, China}

\begin{abstract}
It is demonstrated that the well-regularized hypergeometric functions can be
evaluated directly and numerically. The package {\tt NumExp} is presented for
expanding hypergeometric functions and/or other transcendental functions in a
small regularization parameter. The hypergeometric function is expressed as a
Laurent series in the regularization parameter and the coefficients are
evaluated numerically by using the multi-precision finite difference method. This
elaborate expansion method works for a wide variety of hypergeometric
functions, which are needed in the context of dimensional regularization for
loop integrals. The divergent and finite parts can be extracted from the final
result easily and simultaneously. In addition, there is almost no restriction
on the parameters of hypergeometric functions.
\end{abstract}

\end{frontmatter}

\newpage\noindent
{\bf Program summary}
\begin{description}
\item[]{\em Program title:} {\tt NumExp}
\item[]{\em Catalogue identifier:}
\item[]{\em Program summary URL:}
\item[]{\em Program obtainable from:}
\item[]{\em Licensing provisions:} none
\item[]{\em No.~of lines in distributed program, including test data, etc.:} 1\,682
\item[]{\em No.~of bytes in distributed program, including test data, etc.:} 61\,128
\item[]{\em Distribution format:} {\tt tar.gz}
\item[]{\em Programming language:} Mathematica and/or Python
\item[]{\em Computer:} Any computer where Mathematica or Python is running.
\item[]{\em Operating system:} Linux, Windows
\item[]{\em External routines:} {\tt mpmath} library (for Python)
\item[]{\em Classification:} 4.4, 5, 11.1
\item[]{\em Nature of problem:}
Expansion of hypergeometric functions and/or other transcendental functions in
a small parameter $\epsilon$. These expansions are needed in the context of
dimensional regularization for loop integrals.
\item[]{\em Solution method:}
The hypergeometric function is expressed as a Laurent series in the
regularization parameter $\epsilon$, where the coefficients are evaluated
numerically by the multi-precision finite difference method.
\item[]{\em Restrictions:}
The calculation may be inefficient if the arguments of hypergeometric
functions are close to the convergent boundaries.
\item[]{\em Running time:}
Generally it is below a few seconds, depending on the complexity of the problem.
\end{description}
\newpage

\section{Introduction}

Hypergeometric functions and their extensions are used frequently in the
calculation of Feynman integrals in quantum field theory
\cite{Boos:1990rg,Davydychev:1990cq,Davydychev:1992mt,Davydychev:2000na,
Tarasov:1996br,
Anastasiou:1999cx,Anastasiou:1999ui,Anastasiou:1999bn, Gehrmann:1999as,
Fleischer:2003rm,Weinzierl:2003fx, DelDuca:2009ac}. The hypergeometric
representation makes the calculation of Feynman integrals less tricky and
systematic. In the past decades, several methods, such as the Mellin-Barnes
techniques \cite{Boos:1990rg,Davydychev:1990cq, Smirnov:1999gc,Tausk:1999vh,
Kalmykov:2012rr}, the negative dimensional integration method (NDIM)
\cite{Halliday:1987an, Suzuki:1997yz,Suzuki:1998qv,
Anastasiou:1999ui} or its optimized version, the method of brackets
\cite{Gonzalez:2007ry,Gonzalez:2008xm,Gonzalez:2011nq,Gonzalez:2010nm}, the
differential equations techniques \cite{Kotikov:1990kg}, the dimensional
recurrence relations
\cite{Tarasov:1996br,Fleischer:2003rm, Tarasov:2006nk, Lee:2009dh,
Kniehl:2010aj} and so on, have been used to obtain the
hypergeometric representation of the Feynman integrals. The hypergeometric
representation keeps the expression of the result in a compact form and
spurious singularities cancel each other out automatically. Besides, the
hypergeometric functions have good algebraic and analytic properties, which
make it easy to study the physical problems in different kinematic regions.

In the context of dimensional regularization
\cite{'tHooft:1972fi,Bollini:1972ui}, the
space-time dimension $D=4-2\epsilon$ appears in the parameters of
hypergeometric functions, where the parameter $\epsilon$ regulates infrared
and/or ultraviolet divergences. Formally, the solution in the form of
hypergeometric functions can be expressed as a Laurent series in $\epsilon$,
but in practice this $\epsilon$-expansion of hypergeometric functions is still
not a trivial task.

Recently, a number of pioneering works and elegant algorithms
\cite{Kalmykov:2000qe, Jegerlehner:2002em,Jegerlehner:2003py,
Davydychev:2003mv, Weinzierl:2004bn,
Kalmykov:2006hu,Kalmykov:2006pu,Kalmykov:2007dk,Kalmykov:2008ge} devoted to
construction or explicit evaluation of coefficients of analytic epsilon
expansion of hypergeometric functions around integer, half-integer and
rational values of parameters. Moreover, some packages like
\verb|nestedsums| \cite{Moch:2001zr,Weinzierl:2002hv},
\verb|Xsummer| \cite{Moch:2005uc},
\verb|HypExp| \cite{Huber:2005yg,Huber:2007dx} and
\verb|HYPERDIRE| \cite{Kalmykov:2007pf,Bytev:2011ks}
were developed to perform or provide helps for the analytic
$\epsilon$-expansion of hypergeometric or transcendental functions. These
implementations, however, are restricted to the expansion of some special
classes of hypergeometric functions about integer and/or half-integer
parameters. It is increasingly obvious that more general and convenient
algorithms are needed to meet the demand of practical calculations.

In this work, a numerical algorithm is developed for the $\epsilon$-expansion
of a wide variety of hypergeometric functions. Specifically, the generalized
hypergeometric functions $\HypPFQ{p}{q}$, the Appell hypergeometric functions
and the Horn-type hypergeometric functions of two variables, can all be
expanded in $\epsilon$ by this method. In principle hypergeometric functions
are treated as mathematical objects and there is no restriction on the form of
their parameters. Analytic continuation is performed automatically,
which significantly simplifies the practical calculation. In addition, the
regularized hypergeometric function can be evaluated numerically and lengthy
expressions of the analytic expansion in terms of simpler functions are
avoided. Furthermore, no knowledge of
harmonic or multiple polylogarithms \cite{Vermaseren:1998uu,Remiddi:1999ew,
Gehrmann:2001pz,Gehrmann:2001jv, Vollinga:2004sn, Maitre:2005uu,Maitre:2007kp,
Bonciani:2010ms, Buehler:2011ev} or other
newly defined special functions is needed. As a result, the well regularized
hypergeometric functions can be treated as common functions in the practical
calculations.

The numerical algorithm is based on the ansatz that the regularized
hypergeometric function can be expressed as a Laurent series in the
regularization parameter, where the coefficients are evaluated numerically by
multi-precision finite difference method \cite{Fornberg1988}. Technically, a
finite small numerical parameter $\epsilon_h$ is introduced and the function is
evaluated at a sequence of points $\epsilon_j=(j-\tfrac{1}{2})\epsilon_h$,
then the coefficients are approximated to order $\mathcal{O}(\epsilon_h^{n})$
by optimized finite differences. In principle, there is no restriction on the
order of the expansion. Obviously, parallel computation can be used
to accelerate the numerical $\epsilon$-expansion of hypergeometric functions
because function evaluations can be performed independently.

In the analytic $\epsilon$-expansion method, the coefficients of different
orders are calculated separately. Usually, the expansions of $A(\epsilon)$ and
$B(\epsilon)$ are needed to obtain the expansion of $A(\epsilon)B(\epsilon)$.
Somewhat differently, in the numerical $\epsilon$-expansion method, all
coefficients of the expansion are extracted simultaneously. In addition,
$A(\epsilon)B(\epsilon)$ is expanded directly and there is no need to
calculate the expansions of $A(\epsilon)$ and $B(\epsilon)$ separately. In
other words, the numerical $\epsilon$-expansion method makes the expansion of
complex expression fast and convenient.

The implementations of the algorithm are presented in the form of a
Mathematica package \verb|NumExp.m| and a Python package \verb|numexp.py|.
Note that the \verb|mpmath| \cite{mpmath} package is required by the Python
interface. The algorithm may also be implemented in other computation systems.
Recently the \verb|NumExp| package has been applied successfully to the doubly
heavy hadron spectral density calculation \cite{Huang:2012ti}, where
$\HypPFQ{q+1}{q}$ and/or Appell function $F_4$ are involved.

The calculation may be inefficient if the arguments of hypergeometric
functions are close to the convergent boundaries. Convergence acceleration is
used on and close to the convergent boundaries. However, the convergence
acceleration may degenerate in some cases. This is an intrinsic drawback of
the numerical evaluation of hypergeometric functions. 

The paper is organized as follows. In the next section the theoretical
background and the design of the program are described. Then the usages of the
packages are shown in Sec.~\ref{sec:usage}. Whereafter, some examples are
presented in Sec.~\ref{sec:apps}. Convergence of the algorithm near the
convergent boundaries of hypergeometric functions is discussed in
Sec.~\ref{sec:converg}. Finally, a summary is given in Sec.~\ref{sec:con}.

\section{Theoretical background}
\label{sec:theory}
\subsection{Laurent expansion}
\label{sec:laurent}

Formally, the Laurent expansion of the $\epsilon$-regularized hypergeometric
function can be expressed as
\begin{equation}
\label{eq:lau01}
f(z,\epsilon)=\sum_{n=k}^{\infty} f_n(z) \epsilon^{n},\quad k\in\mathbb{Z},
\end{equation}
where $f(z,\epsilon)$ is a short form of the hypergeometric function,
$z$ the argument vector, and $\epsilon$ the regularization parameter,
respectively. $\epsilon$ can be positive, negative or even complex.
Practically, the new function
$F(z,\epsilon)\equiv\epsilon^{-k}f(z,\epsilon)$ instead of $f(z,\epsilon)$ is
used to perform the Laurent expansion. If $k<0$, this procedure makes the new
function finite for $\epsilon\to0$. On the other hand, if $k>0$, this
procedure helps to increase the precision of the results. So, the following
Laurent expansion
\begin{equation}
\label{eq:lau02}
F(z,\epsilon)=\sum_{n=-m}^{-1}\epsilon^{n}\cdot 0
 +\sum_{n=0}^{\infty} F_n(z) \epsilon^{n}
\end{equation}
will be used to derive the formulas of numerical epsilon expansion of
hypergeometric functions. Note that the spuriously divergent terms
$\sum_{n=-m}^{-1}\epsilon^{n}\cdot 0$ vanish for analytic calculation, but
multi-precision computation is needed to get rid of cancellation errors from such
terms for numerical calculation.

If $f(z,\epsilon_1,\epsilon_2)$ contains more than one regularization
parameters, one can perform the Laurent expansions successively with respect
to $\epsilon_2$, then $\epsilon_1$.

The coefficients $F_n(z)$ are related to the partial differentials of
$F(z,\epsilon)$ with respect to $\epsilon$. In this work, the finite
differences of $F(z,\epsilon)$ will be used for the numerical calculation.

\subsection{Finite difference method}
\label{sec:fdc}

When the regularization parameter $\epsilon$ is set to zero, function
$F(z,\epsilon)$ is usually ill-defined. With non-positive integer parameter,
Gamma function and hypergeometric function contain singularities and may not
give desired results. Moreover, not only the coefficient at the leading order
in $\epsilon$ is needed to be extracted. Therefore, some difference formulas
like
\begin{align*}
F_0(z)&=F(z,\epsilon)+\mathcal{O}(\epsilon),\\
F_1(z)&=\frac{F(z,2\epsilon)-F(z,\epsilon)}{\epsilon}+\mathcal{O}(\epsilon),\\
F_2(z)&=\frac{F(z,3\epsilon)-2F(z,2\epsilon)+F(z,\epsilon)}{2!\,\epsilon^2}
  +\mathcal{O}(\epsilon),
\end{align*}
and so on can be used to extract the coefficients of the Laurent expansion.
That is, $F(z,0)$ will not be used in the calculation.

The precision of the formulas above is very low. Actually, there are many ways
to improve the precision and keep the computational complexity unchanged
\cite{Fornberg1988}. In this work, the algorithm
\begin{equation}
\label{eq:fdcexp}
F_i(z)=\sum_{j=0}^{n}\frac{C_{ij}^{n}}{n!}\,
  F\left(z,\left(j-\tfrac{1}{2}\right)\epsilon_h\right)
  +\mathcal{O}(\epsilon_h^{n-i+1})
\end{equation}
is used for the numerical epsilon expansion. $C_{ij}^{n}$ is the $n$th degree
weight array of the finite difference method, and $0\leq i,j\leq n$. It is
trivial to work out these constant arrays by solving some linear equations.
For example,
\begin{equation}
\label{eq:fdc2}
C_{ij}^{2}=
\left(\begin{array}{rrr}
\tfrac{3}{8} & \tfrac{3}{4} & -\tfrac{1}{8} \\
-1 & 1 & 0 \\
1 & -2 & 1
\end{array}\right).
\end{equation}

The computation of $F(z,(j-\tfrac{1}{2})\epsilon_h)$ is the most
time-consuming part of the whole numerical calculation. It is worth noting
that only $n+1$ times of function evaluation is needed to obtain all $F_i(z)$
($i=0,\ldots,n$) to order $\mathcal{O}(\epsilon_h^{n-i})$. In particular,
with $3$ times of function evaluation, one obtains $F_0(z)$ to order
$\mathcal{O}(\epsilon_h^2)$, $F_1(z)$ to order $\mathcal{O}(\epsilon_h^1)$ and
$F_2(z)$ to order $\mathcal{O}(\epsilon_h^0)$, respectively. If $\epsilon_h$ is
numerically small enough and/or $n$ is large enough, one could get the numerical
$\epsilon$-expansion of $F(z,\epsilon)$ to the desired precision.

\subsection{Precision}
\label{sec:prec}

However, $\epsilon_h$ cannot be too small. From Eq.~\eqref{eq:lau02} and
Eq.~\eqref{eq:fdcexp}, it is easy to see that $-(m+n+1)\lg(|\epsilon_h|)$
digits of working precision is needed to obtain $F_i(z)$ to order
$\mathcal{O}(\epsilon_h^{n-i})$. If $\epsilon_h$ is too small and the
working precision is insufficient, the precision of the low order coefficients
is restricted by the working precision and the high order coefficients will be
inaccurate or meaningless. On the other hand, higher precision means lower
speed of computation. Therefore, one has to find a balance between precision
and efficiency.

With a given working precision, one can use either smaller $\epsilon_h$ (and
less function evaluations) or larger $\epsilon_h$ (and more function
evaluations) to obtain the coefficients of the numerical $\epsilon$-expansion.
For example, with $15$ digits of working precision and $\epsilon_h=10^{-5}$,
$F_0(z)$, $F_1(z)$ and $F_2(z)$ have roughly $15$, $10$ and $5$ digits of
precision, respectively. If $\epsilon_h=10^{-3}$, $F_0(z)$, $F_1(z)$,
$F_2(z)$, $F_3(z)$ and $F_4(z)$ have roughly $15$, $12$, $9$, $6$ and $3$
digits of precision, respectively. It is quite obvious that lower coefficients
can be evaluated to high precision even with a large $\epsilon_h$ at the cost
of more function evaluations. It also shows that the precision of the
coefficients can be increased in a trivial way.

Multi-precision computation is the key to the success of the numerical
$\epsilon$-expansion method. Mathematica and the Python library \verb|mpmath|
\cite{mpmath} provide algorithms to evaluate hypergeometric functions
$\HypPFQ{p}{q}$ and lots of special functions to arbitrary numerical
precision. In particular \verb|mpmath| provides a function \verb|hyper2d()| to
evaluate a wide variety of hypergeometric functions of two variables, such as
Appell functions $F_1$, $F_2$, $F_3$ and $F_4$, which occur in the
calculations involving massive particles
\cite{Davydychev:1992mt,Shpot:2007bz,Kniehl:2011ym,Huang:2012ti}.
With the help of these computation systems, the package \verb|NumExp| can be
used to perform $\epsilon$-expansion of hypergeometric functions in a
numerical and efficient way.

\section{Usage}
\label{sec:usage}

The algorithm in the preceding section is implemented in the package
\verb|NumExp|, and the package can be obtained from the ancillary files of
\eprint. The package provides both Mathematica and Python interfaces for users.

\subsection{Mathematica interface}

After a successful installation, the Mathematica package \verb|NumExp.m| may be
loaded with the command
\begin{verbatim}
    In[1]:= << NumExp`
\end{verbatim}
The package provides several public functions:
\begin{itemize}
\item \verb|NumExpFDC[n]|
generates the $n$th degree weight array $C_{ij}^{n}$. If $n=2$, the array in
Eq.~\eqref{eq:fdc2} will be generated. In most cases, this function will not
be used directly.

\item \verb|NumExp[expr,ep,eh,n]| or \verb|NumExp[expr,{ep,eh,n}]|
is the main function of this package. It expands \verb|expr| to order
$\mathcal{O}({\tt ep}^{n})$. {\tt eh} is the numerical step parameter in
Eq.~\eqref{eq:fdcexp}. The usage of this function is illustrated by the
following example:
\begin{verbatim}
    In[2]:= f[e_,z_]:=Hypergeometric2F1[1,-e,1-e,z];
    In[3]:= NumExp[f[e,3],e,10^(-4),4]
    Out[3]= {1.0000000000000000000`20-6.3*10^(-21)*I,
       0.69314718055994`16+3.141592653589793`16*I,
      -2.32018042335`12   +3.45139229523`12*I,
      -3.7421220`8        +1.8958710`8*I,
      -3.751`4            +0.6944`4*I}
\end{verbatim}
The calculation is performed with $-(n+1)\lg(|\epsilon_h|)$ digits of working
precision and
the precision of the output is estimated by Mathematica automatically. Note
that the precision of the high order coefficients is slightly overestimated,
which is the effect of the $\mathcal{O}(\epsilon_h^{n+1})$ remainder terms.
One can use smaller $\epsilon_h$ or take higher order expansion to increase
the precision of the coefficients.

With the option: \verb|AddPrecision->np|, additional \verb|np| digits of
precision is added to the working precision of \verb|NumExp|. It is helpful
for minimizing the effect of the spurious divergences in Eq.~\eqref{eq:lau02}. 

Besides, the precision of the input parameters must not be lower than the
working precision of \verb|NumExp|, or the precision of the result will be
limited to the lowest precision of the input parameters.

\item \verb|NumExp[expr,{ep1,eh1,n1},{ep2,eh2,n2}]|
successively find numerical epsilon expansions with respect to \verb|ep2|,
then \verb|ep1|. Explicitly,
\begin{equation*}
f(\epsilon_1,\epsilon_2)=
\begin{pmatrix}
1 \\ \epsilon_1 \\ \epsilon_1^2 \\ \vdots
\end{pmatrix}^T
\begin{pmatrix}
f_{00} & f_{01} & f_{02} & \cdots \\
f_{10} & f_{11} & f_{12} & \cdots \\
f_{20} & f_{21} & f_{22} & \cdots \\
\vdots & \vdots & \vdots & \ddots
\end{pmatrix}
\begin{pmatrix}
1 \\ \epsilon_2 \\ \epsilon_2^2 \\ \vdots
\end{pmatrix}.
\end{equation*}
\verb|NumExp| is called twice to obtain the coefficient array $\{f_{ij}\}$ and
the precisions of these two expansions are assigned automatically. Usually,
only the coefficients $\{f_{0j}\}$ in the first row is of interest to users.
The evaluation of the function is performed with
$-(n_1+1)\lg(|\epsilon_{h1}|)-(n_2+1)\lg(|\epsilon_{h2}|)$ digits of working
precision. In addition, the option \verb|AddPrecision| can be assigned to two
\verb|NumExp| expansions separately and explicitly. Then additional digits of
precision will be added to the working precision. This function can be used,
for example, in the following way
\begin{verbatim}
    NumExp[Hypergeometric2F1[1+ep2,-e,1-e,3],
      {ep2,10^-5,2,AddPrecision->5},{e,10^-4,4},AddPrecision->10]
\end{verbatim}
and it may be useful if auxiliary regularization parameter is used in the
calculation (see Eq.~\eqref{eq:pipert} and Ref.~\cite{DelDuca:2009ac}).
It is possible and easy to extend the method to expand functions
with more regularization parameters.

\item \verb|ListNumExp[lst,eh,n]| or \verb|ListNumExp[lst,{eh,n}]|
extracts the coefficients of the $\epsilon$-expansion from a list of values
$f(e_0)$, $f(e_1)$, $\ldots$, $f(e_m)$ which are evaluated beforehand with
$-(n+1)\lg(|\epsilon_h|)$ digits of working precision and the point
$e_j=(j-\tfrac{1}{2})\epsilon_h$.
\verb|eh| is the numerical step parameter and $n\leq m$. The code
\begin{verbatim}
    In[4]:= lst=Table[f[(j-1/2)*10^(-4),3],{j,0,6}];
    In[5]:= ListNumExp[lst,10^(-4),4]
\end{verbatim}
will generate the same results. If $n<m$, the first $n+1$ (from $1$ to $n+1$)
values of \verb|lst| will be used by \verb|ListNumExp|. Note that the step
parameter \verb|eh| must be consistent with the parameter in the $f(e_j)$
evaluations, or the expansion will be nonsense.

This function is especially useful because the time-consuming $f(e_j)$
evaluations can be performed independently, which makes the parallel
computation possible.
Besides, one can also use other programs or packages (like \verb|mpmath|
\cite{mpmath}) to evaluate some special functions unknown to Mathematica and
then use \verb|ListNumExp| to obtain the epsilon expansion.

\item \verb|ListNumExp[ary,{eh1,n1},{eh2,n2}]|
successively find numerical epsilon expansions of the array \verb|ary| with
respect to \verb|ep2|, then \verb|ep1|. The array $\{f(e_{1i},e_{2j})\}$ is
evaluated beforehand with
$-(n_1+1)\lg(|\epsilon_{h1}|)-(n_2+1)\lg(|\epsilon_{h2}|)$ digits of precision
and the point
$(e_{1i},e_{2j})=((i-\tfrac{1}{2})\epsilon_{h1},(j-\tfrac{1}{2})\epsilon_{h2})$.
\end{itemize}

\subsection{Python interface}

Since the Python library \verb|mpmath| \cite{mpmath} provides an extensive set
of transcendental functions and these functions can be evaluated to high
precision in an efficient manner, it is worth to develop a Python interface of
the numerical epsilon expansion algorithm.

The Python package \verb|numexp.py| is specially designed for numerical
$\epsilon$-expansion of hypergeometric functions. Since $\epsilon$ is used to
regularize the expression, these functions may contain spurious divergences at
$\epsilon=0$ and should not be evaluated at $\epsilon=0$ directly. In the
package, the function is evaluated at a list of points
$(j-\tfrac{1}{2})\epsilon_h$ and the coefficients are evaluated by finite
differences of these values.

It is worth noting that \verb|mpmath| \cite{mpmath} provides a function
\verb|taylor()| to produce a degree-$n$ Taylor polynomial around the point $x$
of the given function $f(x)$. The differentials are approximated by finite
differences and this function can handle singularity of $f(0)$ by shifting the
points half of a step length $h/2$. However, \verb|taylor()| is pretty slow
when the precision and the order of expansion are high because the default
step parameter $h$ is related to the working precision and is too small. If a
fairly large $h$ is specified, the speed increases while the precision
decreases. Moreover, if the \verb|singular| option is set to \verb|True|,
which is needed by the expansion of regularized hypergeometric functions,
order $\mathcal{O}(n^2)$ instead of order $\mathcal{O}(n)$ algorithm is used by
\verb|taylor()| and the calculation is roughly $n$ times slower.

\verb|numexp.py| always uses $\mathcal{O}(n)$ algorithm to calculate the
coefficients of the expansion. With the given number of function evaluations,
every coefficient is calculated to the best precision. The error is controlled
in a systematic way and the precision is guaranteed as mentioned in
Sec.~\ref{sec:prec}. Therefore, \verb|numexp.py| is recommended for numerical
$\epsilon$-expansion of the regularized hypergeometric functions.

\verb|numexp.py| provides two public functions:
\begin{itemize}
\item \verb|numexp(f,eh,n,args=[])|
expands \verb|f(ep,*args)| to order $\mathcal{O}({\tt ep}^{n})$. \verb|args| is
the argument list of function $f$ and {\tt eh} is the numerical step
parameter. The usage of this function is illustrated by the following example:
\begin{verbatim}
    In[1]: from numexp import *
    In[2]: mp.dps = 20
    In[3]: f2 = lambda ep: hyp2f1(1,-ep,1-ep,3)
    In[4]: numexp(f2,mpf('1e-4'),4)
    Out[4]:
    [(1.0000000000000000001-6.2579670696103073837e-21j),
    (0.69314718055994379081+3.1415926535897933231j),
    (-2.320180423351603576 +3.451392295225348322j),
    (-3.7421219941535409023+1.8958709608982570633j),
    (-3.7510818776117221586+0.69441875410183733333j)]
\end{verbatim}
These five coefficients have roughly $20$, $15$, $12$, $7$ and $3$ digits of
precisions. One can take the first $n-1$ coefficients and discard the last one
or two. This may be an appropriate strategy for the $\epsilon$-expansion.

If the function $f()$ contains more than one argument, all other arguments
except $\epsilon$ should be put in a list.
\begin{verbatim}
    In[5]: f3 = lambda ep,a,z: hyp2f1(a,-ep,1-ep,z)
    In[6]: numexp(f3,mpf('1e-4'),4,[1,3])
\end{verbatim}

With the option: \verb|addprec=np|, additional \verb|np| digits of
precision is added to the working precision of \verb|numexp|. It is helpful
for minimizing the effect of the spurious divergences in Eq.~\eqref{eq:lau02}.

\item \verb|listnumexp(lst,eh,n)|
extracts the coefficients of the $\epsilon$-expansion from a list of values
$f(e_0)$, $f(e_1)$, $\ldots$, $f(e_m)$ which are evaluated beforehand. {\tt
eh} is the numerical step parameter and $n\leq m$. If $n<m$, the first $n+1$
(from $0$ to $n$) values of \verb|lst| will be used by default.
\end{itemize}

In the following of this work, only accurate digits of the result will be
kept in the results for the sake of simplicity.

\section{Examples}
\label{sec:apps}

In this section some examples for the numerical $\epsilon$-expansion of
hypergeometric functions are given. So as to keep the expressions short, fairly
large $\epsilon_h$ is used in the follow calculation. If the coefficients of
the $\epsilon$-expansion are needed to be evaluated to a higher precision,
smaller $\epsilon_h$ should be used.

\subsection{Hypergeometric functions and analytic continuation}
\label{sec:eg01}

The simple Gauss hypergeometric function
\begin{equation}
\label{eq:eg01a}
A(z,\epsilon)=\HypergeometricPFQ{2}{1}{\epsilon,\epsilon}{1-2\epsilon}{z}
\end{equation}
can be analytically expanded by several packages
\cite{Weinzierl:2002hv,Huber:2005yg}. Alternatively, it is easy to expand
this function by \verb|NumExp| in a numerical way. For $z=0.3$, one gets
\begin{align}
A(0.3,\epsilon)
=& 0.999999999999999999
  +0.326129510\,\epsilon^2
  +0.737565\,\epsilon^3
  +\mathcal{O}(\epsilon^4) ,
\end{align}
where the input parameters $\epsilon_h=10^{-4}$ and $n=4$ are used and only
accurate digits are kept in the result.

If the hypergeometric function is divergent for $\epsilon\to0$, negative power
terms of $\epsilon$ occur in the Laurent expansion. For the function
\begin{equation}
B(z,\epsilon)=\HypergeometricPFQ{2}{1}{2+\epsilon,1+\epsilon}{-2\epsilon}{z},
\end{equation}
one can use $\epsilon\,B(z,\epsilon)$ instead of $B(z,\epsilon)$ to extract
the coefficients, as stated in Sec.~\ref{sec:laurent}. Explicitly,
\begin{align}
B(0.5,\epsilon)
=& 4.00000000000000000000\,\epsilon^{-1}
  +0.9999999999999999 \cr
 &+6.408403647539\,\epsilon
  -10.5952025\,\epsilon^2
  +\mathcal{O}(\epsilon^3) .
\end{align}

If $|z|>1$, analytic continuation is used to evaluate the hypergeometric
function. Numerically, one gets
\begin{align}
A(3,\epsilon)
=& 0.999999999999999999999
  +(2.320180423313-3.451392295223\,i)\epsilon^2 \cr
 &-(5.64797470+12.27713795\,i)\epsilon^3
  -(30.5240+8.050\,i)\epsilon^4
  +\mathcal{O}(\epsilon^5) ,
\end{align}
where analytic continuation is performed automatically in the numerical
calculation.

It is worth noting that analytic continuation is a non-trivial issue for the
analytic expressions generated by \verb|HypExp| and others
\cite{Maitre:2007kp,Huber:2007dx}. If one directly set $z=3$ in the analytic
expansion of Eq.~\eqref{eq:eg01a}, the coefficient of $\epsilon^4$ term will
be $-30.5241-7.35203\,i$, where the imaginary part is incorrect. Technically,
a small imaginary part should be introduced into the argument to keep it on
the correct side of the branch cut. That is, one can use $z=3-10^{-20}\,i$ to
reproduce the correct coefficient.

There is another issue on the analytic continuation of the hypergeometric
functions. For the $\HypPFQ{q+1}{q}$ functions, the analytic continuation
formula for $|z|>1$ is
\begin{align}
\HypergeometricPFQ{q+1}{q}{a_1,\ldots,a_{q+1}}{b_1,\ldots,b_q}{z}
=&\frac{\Gamma(b_1)\cdots\Gamma(b_q)}{\Gamma(a_1)\cdots\Gamma(a_{q+1})}
  \sum_{i=1}^{q+1}
  \frac{\Gamma(a_i)\prod_{j=1;j\neq i}^{q+1}\Gamma(a_j-a_i)}
                  {\prod_{j=1}^{q}\Gamma(b_j-a_i)} \cr
 &\times
  (-z)^{-a_i}
  \HypergeometricPFQ{q+1}{q}%
    {a_i,\{1+a_i-b_k\}_{k=1,\ldots,q}}%
        {\{1+a_i-a_k\}_{k=1,\ldots,q+1;k\neq i}}%
    {\frac{1}{z}},
  \label{eq:q1fqAC}
\end{align}
where $a_j-a_i\notin\mathbb{Z}$. If $a_j-a_i\in\mathbb{Z}$, one can introduce
another auxiliary regularization parameters $\epsilon'$ into the parameters
$a_i$'s to regularize the analytically continued expression. With $a_i\to
a_i+\alpha_i\epsilon'$ and $\alpha_j-\alpha_i\neq0$, the analytically
continued expression is well regularized. Technically, it is advisable to use
irrational $\alpha_i$'s to fulfill the requirement of
$\alpha_j-\alpha_i\neq0$. Note that the expression is finite in $\epsilon'$
and $\epsilon'$ can be set to a small enough value in the numerical
calculation.

In Ref.~\cite{Huang:2012ti} the doubly heavy hadron spectral density has been
expressed in the form of $\epsilon$-regularized hypergeometric functions. A
part of the correlation function is
\begin{align}
\Pi_1^{\mathrm{pert}}(z)=
\frac{3}{64\pi^6}\biggl\{
 &\frac{\Gamma(-\epsilon-4)\Gamma(-\epsilon-2)^2\Gamma(-\epsilon)}
       {\Gamma(2-\epsilon)\Gamma(-2\epsilon-4)} \cr
 &\times
  \HypergeometricPFQ{3}{2}
    {-\epsilon-4,-\epsilon-2+\sqrt{2}\epsilon',-\epsilon+\sqrt{3}\epsilon'}
    {-\epsilon-\tfrac{3}{2},2-\epsilon}{z} \cr
-&\frac{\Gamma(-\epsilon-4)\Gamma(-\epsilon-2)^2\Gamma(-\epsilon)}
       {\Gamma(3-\epsilon)\Gamma(-2\epsilon-4)} \cr
 &\times
  \HypergeometricPFQ{3}{2}
    {-\epsilon-4,-\epsilon-2+\sqrt{2}\epsilon',-\epsilon+\sqrt{3}\epsilon'}
    {-\epsilon-\tfrac{3}{2},3-\epsilon}{z}
  \biggr\} .
  \label{eq:pipert}
\end{align}
The auxiliary $\epsilon'$ is introduced to regularize the analytically
continued expression. In Ref.~\cite{DelDuca:2009ac}, a similar regulator
$\delta$ was used to cancel out the spurious divergences. By using
\begin{equation}
\label{eq:rhoim}
\rho(z)
=\frac{1}{2\pi}\Im\left\{\Pi(z+i\varepsilon)-\Pi(z-i\varepsilon)\right\}
=-\frac{1}{\pi}\Im\Pi(z-i\varepsilon)
\end{equation}
and setting $\epsilon'=10^{-10}$, it is easy to extract the spectral density
from the $\epsilon$-expansion of Eq.~\eqref{eq:pipert}. One can also use the
two dimensional version of \verb|NumExp| (see Sec.~\ref{sec:usage}) to
obtain the ${\epsilon'}^0$ term to a higher precision. Although the Laurent
expansion of the correlation function $\Pi(z)$ contains $\epsilon^{-m}$ terms,
their coefficients are all real and the imaginary parts are zero. Actually,
one can directly evaluate the imaginary part of Eq.~\eqref{eq:pipert} to order
$\mathcal{O}(\epsilon^0)$ and no numerical $\epsilon$-expansion is needed at
all. Even so, one can use \verb|NumExp| to improve the precision of the
result.

Another unorthodox regularization instead of the dimensional regularization
has also been used in Ref.~\cite{Huang:2012ti} to obtain well regularized
hypergeometric representation of the spectral density $\rho(z)$, where all
coefficients of $\epsilon^{-m}$ terms are exactly zero. Then one can use a
small $\epsilon$ to evaluate the special density directly.

\subsection{Generalized hypergeometric functions}
\label{sec:eg02}

Integral representations of some hypergeometric functions may be used to
perform the $\epsilon$-expansion. In Ref.~\cite{Carter:2010hi}, the program
\verb|SecDec| has been used to evaluate the $\epsilon$-expansion of two
generalized hypergeometric functions
\begin{align}
A(\epsilon)
&=\HypergeometricPFQ{5}{4}
  {\epsilon,-\epsilon,-3\epsilon,-5\epsilon,-7\epsilon}
  {2\epsilon,4\epsilon,6\epsilon,8\epsilon}{\frac{1}{2}},
  \label{eq:eg02a} \\
B(\epsilon)
&=\HypergeometricPFQ{4}{3}
  {-4\epsilon,-\tfrac{1}{2}-\epsilon,-\tfrac{3}{2}-2\epsilon,\tfrac{1}{2}-3\epsilon}
  {-\tfrac{1}{2}+2\epsilon,-\tfrac{1}{2}+4\epsilon,\tfrac{1}{2}+6\epsilon}
  {\frac{1}{2}}.
  \label{eq:eg02b}
\end{align}
$A(\epsilon)$ and $B(\epsilon)$ can be analytically expanded by \verb|HypExp|
\cite{Huber:2005yg,Huber:2007dx}. \verb|SecDec| \cite{Carter:2010hi} takes
several minutes to obtain the coefficients. It is transparent that
multi-dimensional numerical integration is time-consuming and its precision is
low.

The numerical epsilon expansion method might also be applied to the
$\epsilon$-regularized parameter integrals, if the poles are subtracted
properly. However, the precision of the multi-dimensional numerical
integration is low, which makes it difficult to extract high order
coefficients of the epsilon expansion by the finite difference method.

The hypergeometric representation has the advantage in high precision
calculation. \verb|NumExp| takes less than one second to expand this type of
generalized hypergeometric functions to order $\mathcal{O}(\epsilon^{6})$. The
numerical results are
\begin{align}
A(\epsilon)
=& 0.9999999999999999999
  +0.189532432184360\,\epsilon \cr
 &-2.2990427423\,\epsilon^2
  +55.469019\,\epsilon^3
  -1014.39\,\epsilon^4
  +\mathcal{O}(\epsilon^{5}), \\
B(\epsilon)
=& 0.999999999999999999
  -4.27968776167885\,\epsilon \cr
 &-26.6975474079\,\epsilon^2
  +195.87119\,\epsilon^3
  -7313.7\,\epsilon^4
  +\mathcal{O}(\epsilon^{5}),
\end{align}
where $\epsilon_h=10^{-4}$ and $n=6$ are used to perform the numerical
$\epsilon$-expansion. Note that all inaccurate digits of the coefficients are
discarded. The precision of the results can be improved by using smaller
$\epsilon_h$ and/or larger $n$.

The analytic $\epsilon$-expansion of generalized hypergeometric functions
about half-integer and rational parameters was studied in
Refs.~\cite{Weinzierl:2004bn, Kalmykov:2007dk,Kalmykov:2008ge}. So as to
illustrate the capability of \verb|NumExp|, the numerical $\epsilon$-expansion
of a hypergeometric function with rational and/or irrational parameters is
performed here. For example, the function
\begin{equation}
C(z,\epsilon)
=\HypergeometricPFQ{4}{3}
  {-4\epsilon,-\tfrac{1}{2}-\epsilon,-\tfrac{\pi}{2}-2\epsilon,\tfrac{1}{3}-3\epsilon}
  {-\pi+2\epsilon,-\tfrac{1}{4}+4\epsilon,\tfrac{1}{2}+6\epsilon}
  {z}
\end{equation}
cannot be expanded by \verb|HypExp| or other analytic $\epsilon$-expansion
algorithms, while \verb|SecDec| may be able to expand this function by using
its integral representation. It is not surprising that \verb|NumExp| can
handle this kind of function with no difficulty. Explicitly,
\begin{align}
C(0.5,\epsilon)
=& 1.0000000000000000000
  -1.44555526747927\,\epsilon \cr
 &+3.9383879447\,\epsilon^2
  -266.94735\,\epsilon^3
  +298.66\,\epsilon^4
  +\mathcal{O}(\epsilon^5) ,
\end{align}
where $\epsilon_h=10^{-4}$ and $n=6$ are used in the expansion.

\subsection{Hypergeometric functions of two variables}
\label{sec:eg04}

Hypergeometric functions of two variables occur in the Feynman diagram related
calculations of three kinematic variables and/or masses
\cite{Davydychev:1992mt,Shpot:2007bz,Alkofer:2008dt,Kniehl:2011ym,Huang:2012ti}.
Recently, a few algorithms or packages
\cite{Moch:2001zr,Weinzierl:2002hv,Moch:2005uc,Bytev:2011ks} have been
developed, which can be used to perform the analytic $\epsilon$-expansion of
some special types of hypergeometric functions of two variables. To the
authors' knowledge, these implementations can only be applied to very limited
cases.

The Python library \verb|mpmath| \cite{mpmath} can evaluate any of the $34$
distinct convergent second-order Horn-type hypergeometric series. With the
help of \verb|mpmath|, \verb|NumExp| can be used to perform the numerical
$\epsilon$-expansion of a wide variety of hypergeometric functions of two
variables.

A numerical scheme was provided for the evaluation of Appell $F_1$ of complex
parameters and real arguments in Refs.~\cite{Colavecchia2001,Colavecchia2004}.
Mathematica provides built-in function \verb|AppellF1| for evaluating $F_1$
with general parameters and arguments to arbitrary numerical precision.
Practically, one can also define other Horn-type hypergeometric functions for
Mathematica according to the same algorithm used by \verb|mpmath|. For
instance, the naive definitions of \verb|AppellF2| and \verb|AppellF4| are
presented in the example file of the package. These functions have been used
to check the correctness of the numerical epsilon expansion.

The analytic $\epsilon$-expansion of the Appell function
\begin{equation}
A(x,y,\epsilon)=\AppellF{2}{1,1,\epsilon}{1+\epsilon,1-\epsilon}{x,y}
\end{equation}
is presented in Ref.~\cite{Moch:2001zr,Weinzierl:2002hv}. Here \verb|NumExp|
is used to show the possibility of the numerical $\epsilon$-expansion of such
kind of functions. Explicitly,
\begin{align}
A(0.3,0.4,\epsilon)
=& 1.428571428571428571
  +0.700889880640673\,\epsilon \cr
 &+1.6060080586\,\epsilon^2
  +1.418379\,\epsilon^3
  +\mathcal{O}(\epsilon^4) ,\\
A(0.3,3.4,\epsilon)
=& 1.428571428571428571
  -(2.43800237269678+4.48798950512827\,i)\epsilon \cr
 &-(7.0450867025-6.6246225829\,i)\epsilon^2
  +(6.57967+ 6.095515\,i)\epsilon^3 ,
\end{align}
where $\epsilon_h=10^{-4}$ and $n=4$ are used in the expansion. For
$|x|+|y|>1$, analytic continuation is performed automatically in the numerical
evaluation. The parameters of this Appell function above are pretty simple and
the expression of the analytic expansion is short. If the parameters are quite
complex, the expression will become lengthy and the numerical
$\epsilon$-expansion may be a better choice for the practical calculation.

Presently, the analytic $\epsilon$-expansion of Appell function $F_4$ is valid
only for some specific cases. In Ref.~\cite{DelDuca:2009ac}, $F_4$ was
converted to $\HypPFQ{2}{1}$ or $F_1$, and then \verb|XSummer|
\cite{Moch:2005uc} was used to perform the $\epsilon$-expansion. Practically,
Appell $F_4$ and other hypergeometric functions of two variables can also be
expanded by \verb|NumExp|.

In the doubly heavy hadron spectral density calculation \cite{Huang:2012ti},
Appell $F_4$ occurs if two heavy quarks are different. For the
$(Q\bar{q})_0^*(\bar{Q}'q)_0^*$ molecular state in Ref.~\cite{Zhang:2009vs},
a part of the correlation function can be expressed as
\begin{align}
\Pi^{\mathrm{pert}}(s)
=&\frac{9 m_1^{\epsilon}m_2^{8-\epsilon}}{256\pi^6}
  \Gamma(2\epsilon-4)\Gamma(-\epsilon-2)\Gamma(2-\epsilon)
  \AppellF{4}{2\epsilon-4;-\epsilon-2}{\epsilon-1,2}
    {\frac{m_1^2}{m_2^2},\frac{s}{m_2^2}} \cr
 &+\frac{9 m_1^{4-\epsilon}m_2^{4+\epsilon}}{256\pi^6}
  \Gamma(\epsilon-2)^2 \Gamma(-2\epsilon)
  \AppellF{4}{\epsilon-2;-2\epsilon}{3-\epsilon,2}
    {\frac{m_1^2}{m_2^2},\frac{s}{m_2^2}} ,
  \label{eq:pipertf4}
\end{align}
where the Appell $F_4$ functions are regularized in an unorthodox way.

For $m_1=1$, $m_2=2$ and $s=19$, for example, the prefactors of the first and
the second terms in Eq.~\eqref{eq:pipertf4} can be expanded as
\begin{align}
B_{01}(\epsilon)
=&-10^{-5}\times\left(
   \frac{9.75151}{\epsilon^2}
  +\frac{9.49329}{\epsilon}
  +81.7264
  +76.8361\,\epsilon
  \right)
  +\mathcal{O}(\epsilon^2) , \\
B_{02}(\epsilon)
=&-10^{-5}\times\left(
   \frac{7.31364}{\epsilon^3}
  +\frac{27.0103}{\epsilon^2}
  +\frac{95.1100}{\epsilon}
  +251.522
  \right)
  +\mathcal{O}(\epsilon) ,
\end{align}
respectively. In the same way, the Appell $F_4$ functions of the first and the
second terms in Eq.~\eqref{eq:pipertf4} can be expanded as
\begin{align}
B_{11}(\epsilon)
=&-0.75\,\epsilon^{-1}
  +47.3086
  +(2.27125-6.55331\,i)\epsilon \cr
 &-(93.1352-8.50940\,i)\epsilon^2
  +\mathcal{O}(\epsilon^3) , \\
B_{12}(\epsilon)
=& 1.0
  +1.51042\,\epsilon
  +(17.8362+8.73770\,i)\epsilon^2 \cr
 &+(0.87460-34.6270\,i)\epsilon^3
  +\mathcal{O}(\epsilon^4) ,
\end{align}
respectively. $B_{01}(\epsilon)$ need to be expanded to $\epsilon^1$ because
$B_{11}(\epsilon)$ contains $\epsilon^{-1}$ contribution. Note that analytic
continuation of $F_4$ is performed automatically in the numerical calculation.
Then,
\begin{align}
\Pi^{\mathrm{pert}}(19)
&=B_{01}(\epsilon)B_{11}(\epsilon)+B_{02}(\epsilon)B_{12}(\epsilon) \cr
&=-\frac{0.00492267}{\epsilon^2}
  -\frac{0.00676321}{\epsilon}
  -(0.0380542+0.0000352703\,i) .
\end{align}
It is transparent that the imaginary parts of $\epsilon^{-m}$ terms are all
zero. In fact, one can directly evaluate the imaginary part of
Eq.~\eqref{eq:pipertf4} to order $\mathcal{O}(\epsilon^0)$ and no numerical
$\epsilon$-expansion is needed at all. Even so, \verb|NumExp| can be used to
improve the precision of the result. Moreover, the spectral density $\rho(z)$
can also be expressed by Appell $F_4$ functions and these
$\epsilon$-regularized functions can be evaluated numerically.

\section{Convergence of the algorithm}
\label{sec:converg}

The calculation may be inefficient if the arguments of hypergeometric
functions are close to the convergent boundaries. The standard series
representation may fail to give the value of the hypergeometric function.
Convergence acceleration is used on and close to the convergent boundaries.
However, the convergence acceleration may degenerate in some cases. This is an
intrinsic drawback of the numerical evaluation of hypergeometric functions. 

For the Gauss hypergeometric function $\HypPFQ{2}{1}$, the well-known analytic
continuations or transformations $z\to 1-z$, $z\to z/(z-1)$, $z\to 1/z$, $z\to
1-1/z$ and $z\to 1/(1-z)$ all map $z=e^{\pm i\pi/3}$ into $z=e^{\pm i\pi/3}$.
That is, these two points are always on the convergent boundary, where the
series converges slowly, or not at all. There are recurrence algorithms for
evaluating $\HypPFQ{2}{1}$ in the vicinity of these special points
\cite{Buhring1988,Michel2007}. Mathematica and \verb|mpmath| \cite{mpmath}
have similar strategies to evaluate $\HypPFQ{2}{1}$ near these points and the
speed is fairly fast.

As for the generalized hypergeometric functions $\HypPFQ{q+1}{q}$, the
well-known analytic continuation is $z\to 1/z$ and $|z|=1$ is the convergent
boundary. Convergence acceleration is used on and close to the unit circle.
However, the convergence acceleration may degenerate as $|z-1|\to 0$.
Mathematica can give the result to high precision, though the speed might be a
little slow. Recently, some algorithms for evaluating $\HypPFQ{q+1}{q}$ near
$z=1$ has been presented \cite{Buhring1992,Buhring2001,Buhring2003,
Skorokhodov2005,Bogolubsky2006,Willis2011}. These algorithms may be helpful
for improving and accelerating the evaluation of $\HypPFQ{q+1}{q}$ near $z=1$.

Numerical evaluation of two-variable hypergeometric functions is even
complicated and time-consuming. Presently, only Appell $F_1$ is
well-implemented to some extent. The other Horn-type hypergeometric functions
can be evaluated by rewriting the series so that the inner series has the form
of an ordinary generalized hypergeometric function and thereby can be
evaluated efficiently \cite{mpmath}. The inner generalized hypergeometric
function can be analytically continued to other regions. Consequently, the
two-variable hypergeometric function is analytically continued. Obviously, the
convergent boundaries are complicated. If the variables are close the
convergent boundaries, the series representation may take quite a long time or
even fail to give the correct result. Inevitably \verb|NumExp| will be
inefficient for such cases because \verb|NumExp| highly relies on the
multiple-precision evaluation of hypergeometric functions. In view of the fact
that the analytic epsilon expansion of two-variable hypergeometric functions
is available only for some special cases, \verb|NumExp| can still be very
useful in the practical calculations.

Actually, the users can employ fast algorithms or programs to evaluate the
hypergeometric function, and then use \verb|ListNumExp| or \verb|listnumexp|
of \verb|NumExp| to extract the coefficients of epsilon expansion. The most
time-consuming function evaluations can be performed in a totally parallel
way, which makes the numerical epsilon expansion method of practical uses even
when the hypergeometric function evaluation is pretty slow.

\section{Summary}
\label{sec:con}

The Feynman integrals can be calculated in a less tricky and systematic way by
using the hypergeometric representation. Presently, the analytic
$\epsilon$-expansion methods of hypergeometric functions have been used in
practice. However, the application of such expansion has lots of limitations.
For example, only some classes of hypergeometric functions with the specific
form of parameters can be expanded, and some newly defined special functions
are used as primary elements and the expanded expressions are quite lengthy
which make the physical analysis cumbersome.

In this work it is demonstrated that the well-regularized hypergeometric
functions can be evaluated directly and numerically. An algorithm is developed
and the package \verb|NumExp| is presented for expanding hypergeometric
functions and/or other transcendental functions in a small regularization
parameter. Hypergeometric function is expressed as a Laurent series in the
regularization parameter and the coefficients are evaluated numerically by
using multi-precision finite difference method. This elaborate expansion
method works for a wide variety of (hypergeometric) functions, such as the
generalized hypergeometric functions $\HypPFQ{p}{q}$ (not just the
$\HypPFQ{q+1}{q}$), the Appell
hypergeometric functions and other Horn-type hypergeometric functions of two
variables, which are needed in the context of dimensional regularization for
loop integrals. It is straightforward and easy to obtain the high order
coefficients $F_n(z)$ of the numerical $\epsilon$-expansion.
In addition, there is almost no restriction on the form of parameters of
hypergeometric functions.

The numerical $\epsilon$-expansion method may not be suitable for the
intermediate expressions, but it is good enough for the final results.
Practically, the divergent and finite parts can be extracted from the final
result easily and simultaneously. Moreover, parallel computation can be used
to accelerate the numerical $\epsilon$-expansion of hypergeometric functions.
In other words, the numerical $\epsilon$-expansion method makes the evaluation
of Feynman integrals fast and convenient.

It is possible and easy to extend the method to expand functions with two or
more regularization parameters. It is worth noting that this method can serve
to be an important cross-check of the analytic $\epsilon$-expansion methods
and it can be used in case of the analytic $\epsilon$-expansion is not
available.

\section*{Acknowledgments}

We thank the anonymous referees for providing constructive comments and help
in improving the contents of this paper.
We are also grateful to Mikhail Kalmykov for his technical comments on our work.
We would like to thank Ze-kun Guo for careful reading of the manuscript.
This work is supported by the National Natural Science Foundation
of China under Grant No.~10775105, BEPC National Laboratory Project R\&D and
BES Collaboration Research Foundation, and the project of Wuhan University of
China under the Grant No.~201103013 and 9yw201115.

\appendix
\bibliographystyle{elsarticle-num}
\bibliography{numexp}
\end{document}